\documentclass[a4paper, fleqn]{cas-sc}

\usepackage[square, numbers, sort&compress]{natbib}
\usepackage[figurename=Fig.]{caption}
\usepackage{subfig}
\usepackage{graphicx}
\usepackage{mathrsfs}
\usepackage[scr=esstix, cal=boondox]{mathalfa} % set mathcal for lowercase
\usepackage{tikz}
\usepackage[capitalise]{cleveref}
\usepackage{array, multirow}
\usepackage{makecell}
\usepackage{multirow}

\ExplSyntaxOn %%disable thumbnails
\keys_set:nn
{ stm / mktitle }
{ nologo } %%disable thumbnails
\ExplSyntaxOff %%disable thumbnails

\def\tsc#1{\csdef{#1}{\textsc{\lowercase{#1}}\xspace}}
\tsc{WGM}
\tsc{QE}
\tsc{EP}
\tsc{PMS}
\tsc{BEC}
\tsc{DE}
\begin{document}
  \let\WriteBookmarks\relax \def\floatpagepagefraction{1} \def\textpagefraction{.001}

  % Short title
  \shorttitle{S-NOT}

  % Short author
  \shortauthors{Q. Liu et~al.}

  % Main title of the paper
  \title[mode = title]{Sequential Neural Operator Transformer for
  High-Fidelity Surrogates of Time-Dependent Non-linear Partial Differential Equations}
  % Title footnote mark
  % eg: \tnotemark[1]

  % Title footnote 1.
  % eg: \tnotetext[1]{Title footnote text}
  % \tnotetext[<tnote number>]{<tnote text>}

  % Options: Use if required
  % eg: \author[1,3]{Author Name}[type=editor,
  %       style=chinese,
  %       auid=000,
  %       bioid=1,
  %       prefix=Sir,
  %       orcid=0000-0000-0000-0000,
  %       facebook=<facebook id>,
  %       twitter=<twitter id>,
  %       linkedin=<linkedin id>,
  %       gplus=<gplus id>]
  % \cortext[corresponding]{Corresponding author}
  \author%
  [ncsa,ksu]{Qibang Liu}[orcid=0000-0001-7935-7907] \cormark[1] \ead{qibang@illinois.edu}

  \author%
  [ncsa,mse]{Seid Koric}

  % Address/affiliation
  \affiliation[ncsa]{organization={National Center for Supercomputing Applications, University of Illinois Urbana-Champaign}, city={Urbana}, postcode={61801}, state={IL}, country={USA}}
  \affiliation[mse]{organization={The Grainger College of Engineering, Department of Mechanical Science and Engineering, University of Illinois Urbana-Champaign}, city={Urbana}, postcode={61801}, state={IL}, country={USA}}
  \affiliation[ksu]{organization={Department of Industrial and Manufacturing Systems Engineering, Kansas State University}, city={Manhattan}, postcode={66506}, state={KS}, country={USA}}

  % Here goes the abstract
  \begin{abstract}
    Partial differential equations (PDEs) are fundamental to modeling complex and
    nonlinear physical phenomena, but their numerical solution often requires
    significant computational resources, particularly when a large number of forward
    full solution evaluations are necessary, such as in design, optimization,
    sensitivity analysis, and uncertainty quantification. Recent progress in
    operator learning has enabled surrogate models that efficiently predict full
    PDE solution fields; however, these models often struggle with accuracy and
    robustness when faced with highly nonlinear responses driven by sequential
    input functions. To address these challenges, we propose the Sequential
    Neural Operator Transformer (S-NOT), a architecture that combines gated recurrent
    units (GRUs) with the self-attention mechanism of transformers to address time-dependent,
    nonlinear PDEs. Unlike S-DeepONet (S-DON), which uses a dot product to merge
    encoded outputs from the branch and trunk sub-networks, S-NOT leverages attention
    to better capture intricate dependencies between sequential inputs and spatial
    query points. We benchmark S-NOT on three challenging datasets from real-world
    applications with plastic and thermo-viscoplastic highly nonlinear material
    responses: multiphysics steel solidification, a 3D lug specimen, and a dogbone
    specimen under temporal and path-dependent loadings. The results show that S-NOT
    consistently achieves a higher prediction accuracy than S-DON even for data
    outliers, demonstrating its accuracy and robustness for drastically accelerating
    computational frameworks in scientific and engineering applications.
  \end{abstract}

  % Use if graphical abstract is present
  % \begin{graphicalabstract}
  %   \includegraphics[width=\textwidth]{S-NOT.pdf}
  % \end{graphicalabstract}

  % Research highlights
  % \begin{highlights}
  %   \item A novel Sequential Neural Operator Transformer (S-NOT) is introduced to
  %   efficiently solve time-dependent nonlinear PDEs

  %   \item S-NOT integrates attention mechanisms to captures complex
  %   relationships between input functions and query points, improving generalization
  %   for time-dependent physical phenomena

  %   \item S-NOT significantly outperforms the Sequential DeepONet (S-DeepONet)
  %   in prediction accuracy
  % \end{highlights}

  % Keywords
  % Each keyword is seperated by \sep

  \maketitle

  \section{Introduction}
  Partial differential equations (PDEs) are central to modeling a wide range of physical
  phenomena, as they encode the governing laws of physics. Traditional numerical
  methods—such as finite difference, finite element, and finite volume approaches—have
  been widely used to obtain solutions to these equations. Although these
  methods can achieve high-fidelity results and capture intricate behaviors, they
  are often computationally expensive and time-consuming, especially for large-scale
  or high-dimensional problems. As a result, running such simulations can be
  prohibitive in terms of both time and computational resources.

  Recently, data-driven deep learning approaches—commonly known as surrogate models—have
  emerged to address these computational challenges across diverse domains,
  including solid mechanics \citep{kiranyaz2021exploiting,yang2021deep}, fluid
  mechanics, electromagnetics \citep{sun2021development,peng2022rapid}, acoustics
  \citep{konuk2021physics,borrel2024sound}, and chemical polymerization \citep{liu2024adaptive,cai2024towards},
  among others. Leveraging neural networks, these methods can efficiently approximate
  PDE solutions, enabling substantial reductions in computational cost and
  accelerating simulations while maintaining high accuracy. Such surrogate models
  are particularly beneficial for tasks that require repeated PDE evaluations,
  such as uncertainty quantification \citep{abdar2021review,cheng2023machine},
  sensitivity analysis \citep{li2021modeling,wang2023prediction}, and
  optimization using gradient-based methods \citep{dogo2018comparative,daoud2023gradient}
  or generative AI designes \citep{liu2025univariate,liu2025towards}.

  DeepONet \citep{lu2021learning} is a widely used surrogate model that employs artificial
  neural networks to approximate nonlinear operators mapping input functions to outputs
  at specified query points:
  \begin{equation}
    \label{eq:NOT}G_{\theta}: \mathcal{F}\rightarrow \mathcal{G}(\mathbf{x}),
  \end{equation}
  where $G_{\theta}$ denotes the neural operator parameterized by $\theta$, $\mathcal{F}$
  is the space of input functions, and $\mathcal{G}$ is the space of output
  functions evaluated at query points $\mathbf{x}$.

  The original DeepONet framework \citep{lu2021learning} consists of two fully connected
  neural networks (FNNs): a branch network that encodes the input function at
  specified sensor locations, and a trunk network that encodes the spatiotemporal
  coordinates of the solution. The outputs of these networks are combined via a
  dot product to predict the solution, which is then compared to reference values
  from conventional numerical methods. DeepONet has been successfully applied to
  predict full-field solutions in a range of fields, including materially nonlinear
  solid mechanics \citep{koric2024deep}, fracture mechanics \citep{goswami2022physics},
  aerodynamics \citep{zhao2023learning}, acoustics \citep{xu2023training}, seismology
  \citep{haghighat2024en}, digital twins \citep{kobayashi2024improved}, heat transfer
  \citep{sahin2024deep,koric2023data}, and real-time monitoring \citep{hossain2025virtual},
  among others. However, these applications typically focus on static or stationary
  loads, and do not address time-dependent loading scenarios. In practice, many
  real-world loads—such as wind, vibration, or impact—are inherently time-dependent.
  A standard FNN in the branch network is not designed to preserve the causality
  present in sequential input data. To overcome this limitation, \citet{he2024sequential}
  proposed the Sequential DeepONet (S-DON), which replaces the FNN in the branch
  network with gated recurrent units (GRUs) to better capture the sequential characteristics
  of time-dependent inputs.

  While S-DON has demonstrated notably improved performance for time and path-dependent
  problems compared the original DeepONet, its reliance on a simple dot
  product to merge branch and trunk outputs can limit its ability to capture
  complex dependencies, especially across all challenging scenarios. Recent
  advances have introduced the use of transformer-based attention mechanisms
  \citep{Ashish2017atten} in neural operators, leading to the development of
  Neural Operator Transformers (NOT) \citep{cao2021choose,liu2022ht,li2022transformer,hao2023gnot,liu2025towards,liu2025geometry}.
  Moreover, the recent work by  \citet{shih2025transformers} established a theoretical groundwork that a transformer with cross-attention possesses the universal approximation property as an operator learning model.
  Unlike the simple dot product, the attention mechanism enables each query point
  to selectively aggregate information from the input functions, allowing the
  model to better capture intricate relationships and dependencies. Moreover,
  since transformers were originally designed to model long-range dependencies
  in sequential data, they are particularly well-suited for learning the complex
  spatiotemporal interactions present in PDE solutions under time and path-dependent
  loading.

  In this work, we introduce the Sequential Neural Operator Transformer (S-NOT),
  a novel framework for solving time and path-dependent nonlinear problems with improved
  accuracy and generalization. S-NOT combines the sequential modeling
  capabilities of gated recurrent units (GRUs) with the self-attention and cross-attention
  mechanisms of transformer architectures to effectively encode sequential
  loading data and predict solution fields at query points. We evaluate S-NOT on
  three representative datasets: multiphysics steel solidification, a 3D lug specimen,
  and a dog-bone specimen under time-dependent loading. Comparative results demonstrate
  that S-NOT consistently achieves higher prediction accuracy than S-DON across all
  test cases.

  This manuscript presents the Sequential Neural Operator Transformer (S-NOT) framework,
  detailing its architecture, methodology, and performance. \cref{sec:methodology}
  outlines the S-DON and S-NOT models, while \cref{sec:examples_datasets}
  introduces the three benchmark datasets used for training and evaluation. The main
  results are discussed in \cref{sec:results}, and key findings are summarized
  in \cref{sec:conclusion}.

  \section{Methodology}
  \label{sec:methodology}

  \subsection{Sequential DeepONet}

  \begin{figure}[htbp]
    \centering
    \includegraphics[width=\textwidth]{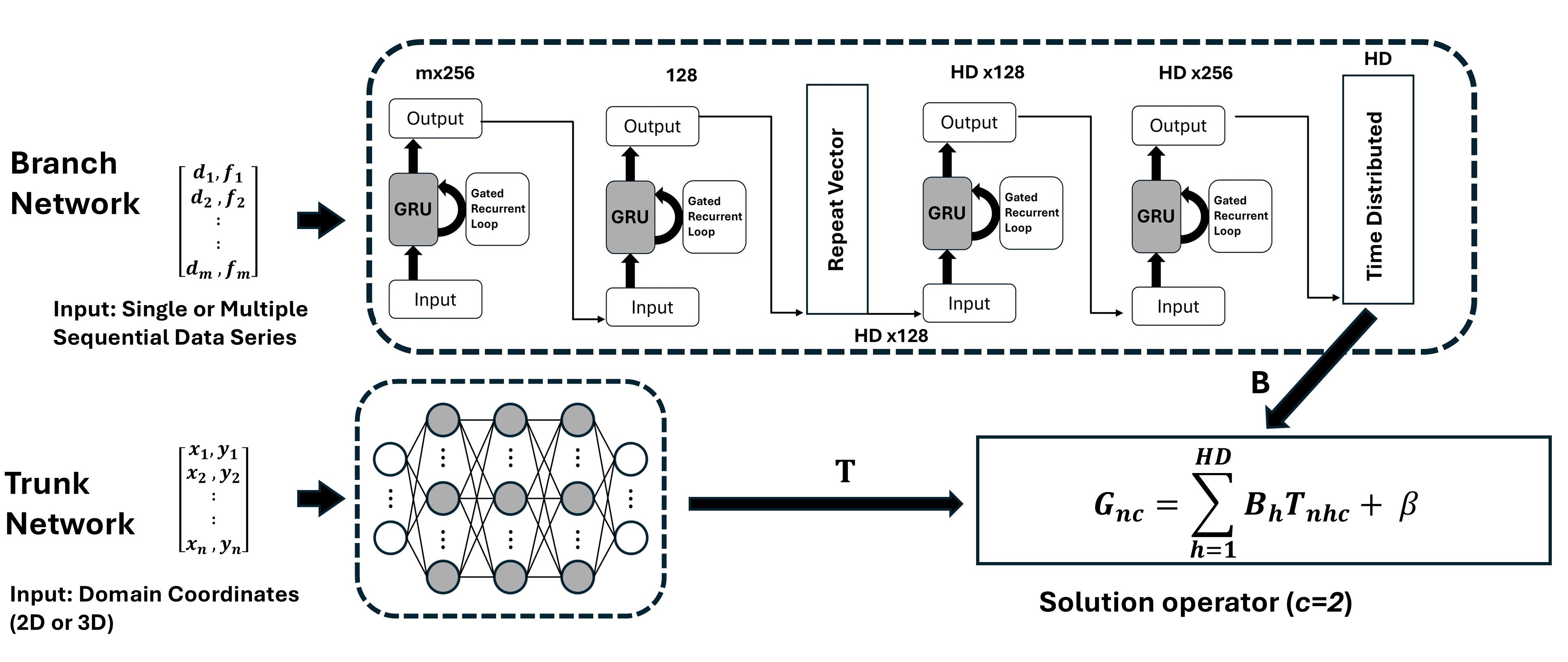}
    \caption{S-DON architecture.}
    \label{fig:sdon}
  \end{figure}

  The original DeepONet framework \citep{lu2021learning} has two fully connected
  neural networks (FNNs): the branch network, which encodes the input function
  at designated sensor locations, and the trunk network, which encodes the
  spatiotemporal coordinates of the solution. DeepONet integrates the outputs of
  these two networks via a dot product to forecast the solution, which is
  subsequently compared with target values generated from conventional numerical
  approaches. The prediction error is measured via a loss function, whose
  gradients with respect to the trainable parameters of neural networks are
  calculated during backpropagation, allowing the optimizer to repeatedly enhance
  the weights and biases in both networks. With adequate training
  epochs—iterations of feedforward and backpropagation—DeepONet acquires the
  ability to forecast the whole solution field throughout the domain.

  In cases characterized by irreversible material deformation or thermo-mechanical
  multiphysics, where plastic or thermo-viscoplastic behavior creates path or time
  dependency, the complete load history is essential for precisely ascertaining
  the final stress state. The Sequential DeepONet (S-DON) \citep{he2024sequential}
  was recently introduced to enhance the management of sequential input
  functions in these scenarios. This architecture substitutes the conventional feedforward
  neural network in the branch network with an encoder-decoder model utilizing gated
  recurrent units (GRUs), better at capturing the sequence of the input data. Gating
  methods in GRUs address vanishing gradient issues and facilitate the learning of
  long-term dependencies, rendering them more effective for sequential learning
  than conventional recurrent neural networks. Although the training duration was
  extended compared to the original DeepONet, the S-DON model substantially
  enhanced predictive accuracy \citep{he2024sequential} and became the standard in
  sequential operator learning. \cref{fig:sdon} illustrates the architecture
  with a single branch that encodes single or multiple sequential inputs (such as
  $d_{i}$ and $f_{i}$) in a coupled fashion via a series of encoder-decoder GRU
  cells, yielding an encoded branch output Bh with hidden dimension. The trunk
  network encodes $n$ input domain 2D or 3D coordinates into its
  multidimensional output $T_{nhc}$, where $c$ denotes the number of output solution
  fields predicted concurrently, such as temperature and stress, or plastic strain
  and stress in the examples in this work. The encoded outputs from the branch
  and trunk are combined by a dot product to get the final S-DON prediction
  $\textbf{G}_{nc}$, incorporating a bias $\beta$ along the hidden dimension.
  For a more detailed description of the S-DON architecture and its applications,
  one can consult \citep{he2024sequential,kushwaha2024advanced}.

  \subsection{Sequential Neural Operator Transformer (S-NOT)}
  \label{sec:snot}

  \begin{figure}[htbp]
    \centering
    \includegraphics[width=\textwidth]{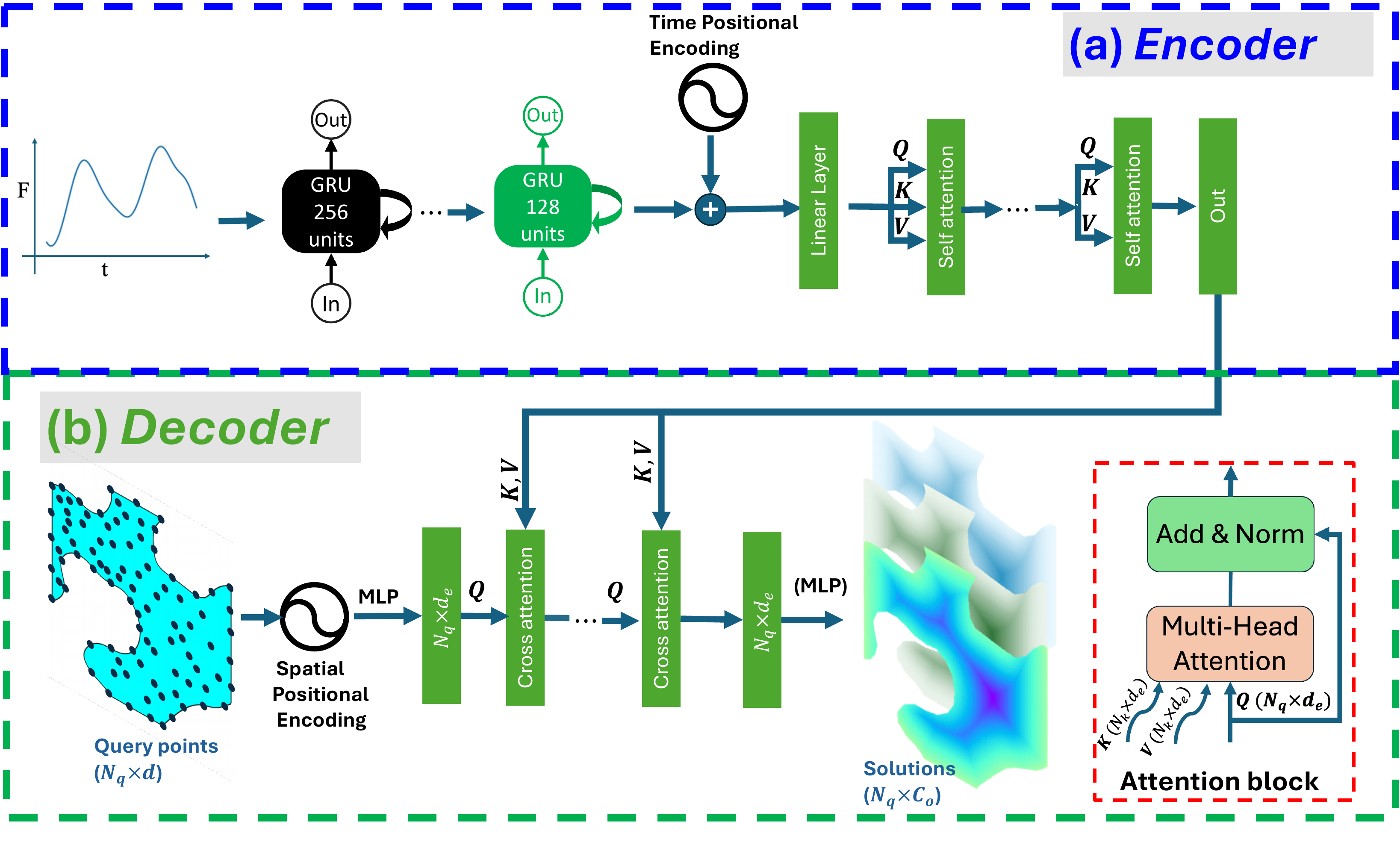}
    \caption{ Overview of the S-NOT architecture.}
    \label{fig:snot}
  \end{figure}

  S-NOT combines the transformer architecture with the neural operator framework
  to predict solution fields at query nodes under varying sequential loading. The
  model consists of two main components: a sequential loading encoder and a
  solution decoder, as shown in \cref{fig:snot}. Both components utilize the
  attention mechanism from the transformer architecture \citep{Ashish2017atten},
  enabling the model to selectively focus on relevant input features. The attention
  operation is defined as
  \begin{equation}
    \text{Attention}(Q, K, V) = \text{softmax}\left(\frac{QK^{T}}{\sqrt{d_{e}}}\right
    )V,
  \end{equation}
  where $Q \in \mathbb{R}^{N_q \times d_{e}}$ is the QUERY matrix, and
  $K, V \in \mathbb{R}^{N_k \times d_{e}}$ are the KEY and VALUE matrices, respectively.
  Here, $d_{e}$ is the embedding dimension, $N_{q}$ is the number of query
  points, and $N_{k}$ is the length of the KEY and VALUE sequences. Each
  attention block incorporates residual connections and layer normalization to
  promote stable training, as illustrated in the lower right of \cref{fig:snot}.

  The encoder processes the sequential loading data using a stack of GRUs, as in
  S-DON, to capture temporal dependencies. The GRU outputs are augmented with
  sinusoidal positional encodings \citep{Ashish2017atten} to incorporate time information
  explicitly. This enriched representation, containing both loading values and temporal
  context, is projected via a linear layer to form the QUERY, KEY, and VALUE
  matrices for multiple self-attention blocks. The self-attention blocks enable the
  model to further selectively focus on the most informative aspects of the sequential
  input functions, improving S-NOT's ability to capture the underlying
  relationships between the elements in input space. The encoder's final output,
  shaped as $N_{k}\times d_{e}$, serves as the KEY and VALUE inputs to the decoder's
  cross-attention layers.

  The solution decoder generates predictions at the query points by leveraging the
  sequential representations from the encoder. Each query point is first
  embedded using a NeRF-style positional encoding \citep{mildenhall2021nerf},
  followed by a multilayer perceptron (MLP), to produce the QUERY matrix of size
  $N_{q}\times d_{e}$, where $N_{q}$ is the number of query points and $d_{e}$
  is the embedding dimension. This QUERY is then combined with the encoder's KEY
  and VALUE outputs through cross-attention blocks, replacing the fixed inner
  product used in S-DON. This approach enables each query point to selectively aggregate
  relevant information from the sequential loading history, thereby enhancing
  the model's capacity to capture complex dependencies and supporting the
  universal approximation property in operator learning. The resulting representations
  of cross-attention are passed through a final MLP to produce the predicted solution
  fields at the query locations. The overall structure of the solution decoder
  is depicted in \cref{fig:snot}(b).

  \section{Example datasets}
  \label{sec:examples_datasets}

  In this section, we introduce the three benchmark datasets used to train and evaluate
  the S-NOT model. The first dataset involves a multiphysics steel
  solidification problem with time-dependent heat flux and displacement boundary
  conditions. The second dataset features a 3D lug specimen subjected to time-varying
  pressure loading. The third dataset consists of a dog-bone specimen under
  sequential displacement loading. These datasets represent a range of challenging,
  time-dependent physical scenarios for assessing the performance and
  generalization of the proposed model.

  \subsection{Multiphysics Steel Solidification}

  \begin{figure}[htbp]
    \centering
    \includegraphics[width=\textwidth]{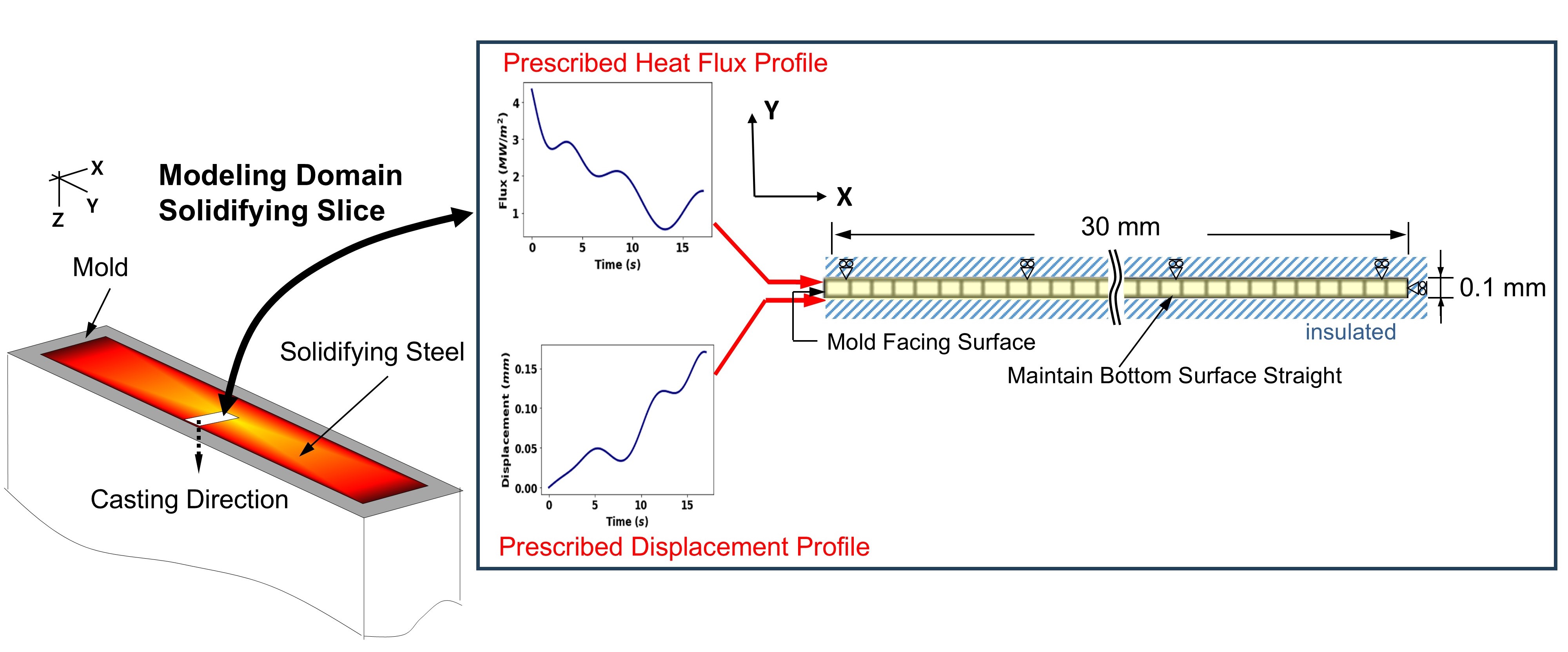}
    \caption{`Slice modeling domain in caster with boundary value conditions.}
    \label{fig:slice_domain_bcs}'
  \end{figure}

  To generate training and validation datasets for our operator learning
  frameworks, in the first example, we use a real-world multiphysics (thermo-mechanical)
  finite element simulation that numerically models the solidification of a 2D
  slice in a continuous steel caster. Over 95\% of steel worldwide is produced
  nowadays by the continuous casting process, which makes this example extremely
  relevant. The solidifying slice is modeled in a Lagrangian reference frame, tracking
  its evolution through the mold region. The thermal evolution is governed by an
  energy conservation equation in \cref{eq:heat_eq} that incorporates material
  density ($\rho$), temperature-dependent conductivity (k), and enthalpy (H), which
  accounts for latent heat due to solid-state transformations such as $\delta$-ferrite
  to austenite transitions.
  \begin{equation}
    \label{eq:heat_eq}\rho\frac{\partial H}{\partial t}= \nabla \cdot (k \nabla T
    )
  \end{equation}

  Given the negligible influence of inertia during solidification, mechanical
  equilibrium is treated as quasi-static in \cref{eq:stress_eq}:
  \begin{equation}
    \label{eq:stress_eq}\nabla \cdot \boldsymbol{\sigma(x)}+\textbf{b}=0
  \end{equation}
  Here, $\sigma$ is the Cauchy stress tensor, and $\textbf{b}$ is the body force
  density. The stress equilibrium is solved concurrently with the thermal field,
  with mechanical behavior captured using a thermo-elastic-viscoplastic formulation.
  The total strain rate tensor $\Dot{\varepsilon}$ is decomposed into elastic, inelastic,
  and thermal contributions; $\Dot{\boldsymbol{\varepsilon}}_{el}$,
  $\Dot{\boldsymbol{\varepsilon}}_{ie}$, $\Dot{\boldsymbol{\varepsilon}}_{th}$ in
  \cref{eq:strain_eq}.
  \begin{equation}
    \label{eq:strain_eq}\Dot{\boldsymbol{\varepsilon}}=\Dot{\boldsymbol{\varepsilon}}
    _{el}+\Dot{\boldsymbol{\varepsilon}}_{ie}+\Dot{\boldsymbol{\varepsilon}}_{th}
  \end{equation}
  The thermal strain depends on the temperature distribution and the temperature-dependent
  thermal expansion coefficient, $\alpha$. Stress and strain rates are related
  in \cref{eq:stress_strain_eq}
  \begin{equation}
    \label{eq:stress_strain_eq}\boldsymbol{\sigma}=\boldsymbol{D}:\left(\Dot{\boldsymbol{\varepsilon}}
    -\alpha \textbf{I}\Dot{T}-\Dot{\boldsymbol{\varepsilon}}_{ie}\right)
  \end{equation}
  In this equation, \textbf{D} is a fourth-order tensor that contains the
  temperature-dependent elastic constants for an isotropic material, and \textbf{I}
  is the identity tensor.

  The mechanical response of solidifying steel is highly nonlinear, exhibiting time-
  and temperature-dependent inelastic deformation. For the austenite phase, we
  use a viscoplastic model originally proposed by \citet{kozlowski1992simple},
  which relates the equivalent inelastic strain rate
  $\Dot{\Bar}{\boldsymbol{\varepsilon}}_{ie}$ ie to Von Mises stress
  $\Bar{\sigma}$(MPa), absolute temperature, and steel grade determined by carbon
  content \%C given in Eq. (5)
  \begin{equation}
    \dot{\bar{\varepsilon}}_{\mathrm{ie}}\left[\sec^{-1}\right]=\mathrm{f}_{\mathrm{C}}
    \left(\bar{\sigma}[M P a]-\mathrm{f}_{1}\bar{\varepsilon}_{\mathrm{ie}}\left
    |\bar{\varepsilon}_{\mathrm{ie}}\right|^{\mathrm{f} 2-1}\right)^{\mathrm{f}
    3}\exp \left(-\frac{\mathrm{Q}}{\mathrm{~T}[\mathrm{~K}]}\right)
  \end{equation}
  where
  \begin{equation}
    \begin{aligned}
       & Q=44.465                                   \\
       & f_{1}=130.5-5.128 \times 10^{-3}T[K]       \\
       & f_{2}=-0.6289+1.114 \times 10^{-3}T[K]     \\
       & f_{3}=8.132-1.54 \times 10^{-3}T[K]        \\
       & f_{c}=46,550+71,400(\% C)+12,000(\% C)^{2}
    \end{aligned}
  \end{equation}

  Q is an activation energy constant, and the empirical functions $f_{1}$,$f_{2}$,$f
  _{3}$,$f_{c}$ vary with absolute temperature and carbon content.

  In contrast, $\delta$-ferrite phase, characterized by lower strength and greater
  creep susceptibility, is described using the Zhu power law model \citep{zhu1996coupled}.
  When the $\delta$-ferrite phase exceeds 10\% of the local microstructure, this
  model governs the material response. In mushy zones above the solidus temperature,
  we assume a nearly negligible yield stress using an elastic-perfectly-plastic
  model.

  These constitutive equations are implemented and locally integrated within a
  user-defined material subroutine (UMAT) in the Abaqus implicit finite element
  analysis code \citep{abaqus2022}. The UMAT computes phase fractions and
  interpolates material properties as functions of temperature for a selected
  low-carbon steel (0.09 wt\%C), with solidus and liquidus temperatures of 1480.0
  °C and 1520.7 °C, respectively. More detailed descriptions of this robust multiphysics
  model are provided in \citep{koric2006efficient,zappulla2020multiphysics}.

  The computational domain traveling down the caster, depicted in \cref{fig:slice_domain_bcs},
  is a 2D slice subjected to generalized plane strain assumptions in the axial
  direction to emulate 3D stress states relevant to long and wide caster
  geometries. The domain consists of 300 coupled thermo-mechanical quadrilateral
  elements and 602 nodes. A fixed bottom boundary condition ensures uniform
  vertical strain, while thermal and displacement profiles are applied on the left
  boundary to simulate interaction with the mold.

  Transient boundary conditions are imposed to reflect the casting process: heat
  fluxes evolve over time due to changing contact conditions, and the displacements
  reflect mold taper effects. These profiles are based on plant measurements and
  are perturbed using Gaussian radial basis interpolation to mimic realistic
  noise and variability observed in operating conditions. This ensures that the generated
  data captures a broad spectrum of physical scenarios encountered during casting.
  More details about these thermal and mechanical input sequences can be found
  at \citep{abueidda2021deep}, where they were used for non-operator deep
  learning.

  5000 data samples were generated utilizing the high-throughput and parallel
  CPU processing capabilities of the Delta high-performance cluster at the
  National Center for Supercomputing Applications (NCSA) \citep{delta2025}. Each
  data sample consists of input sequences, and the final lateral stress and temperature
  solutions at the mold exit are used as labels. They were used to train the
  Sequential DeepONet (S-DON) and the Sequential Neural Operator Transformer (S-NOT)
  models. The complete dataset was divided, with 80\% allocated for training and
  the remaining 20\% allocated for testing. Computational nodes equipped with Nvidia
  H200 GPUs in the DeltaAI \citep{deltaai2025} cluster were used to train both S-DON
  and S-NOT models.

  \subsection{3D-LUG}

  \begin{figure}[htbp]
    \centering
    \includegraphics[width=\textwidth]{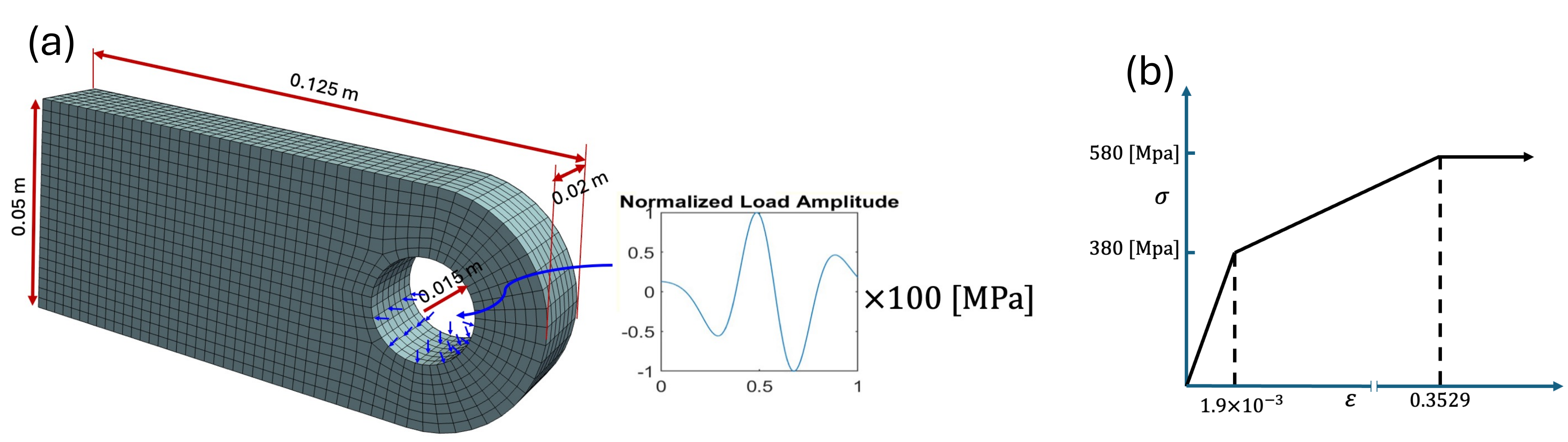}
    \caption{The 3D LUG specimen for FEM simulation. (a) The lug geometry, FE
    mesh, and time-dependent loading profile. (b) the material properties used
    in the simulation.}
    \label{fig:LUG_model}
  \end{figure}

  In the second example, we consider a 3D lug specimen subjected to time-dependent
  pressure loading. The geometry, finite element mesh, and loading profile are
  shown in \cref{fig:LUG_model}(a). Lugs are widely used structural components, and
  accurately predicting their mechanical response is important for assessing system
  performance and reliability. In this setup, a random time-dependent normal pressure
  is applied to the bottom half of the circular hole.

  The lug is discretized using 6,336 second-order hexahedral elements, resulting
  in 29,852 nodes. The material behavior is modeled with an elastic-plastic constitutive
  law with linear isotropic hardening:

  \begin{equation}
    \sigma_{y}(\varepsilon_{p}) = \sigma_{pe}+ H \varepsilon_{p}
  \end{equation}

  where $\sigma_{y}$ is the flow stress, $\varepsilon_{p}$ is the equivalent plastic
  strain, $\sigma_{pe}$ is the initial yield stress, and $H$ is the hardening modulus.
  The specific material properties used are summarized in \cref{fig:LUG_model}(b).

  To generate the dataset, random loading profiles are sampled from a Gaussian
  random field. A total of 10,000 paired samples of loading profiles, the final
  state of von Mises stress, and plastic equivalent strain (PEEQ) are generated.
  The dataset is split into 80\% for training and 20\% for testing.

  \subsection{Dog-bone}

\begin{figure}[htbp]
    \centering
    \includegraphics[width=4in]{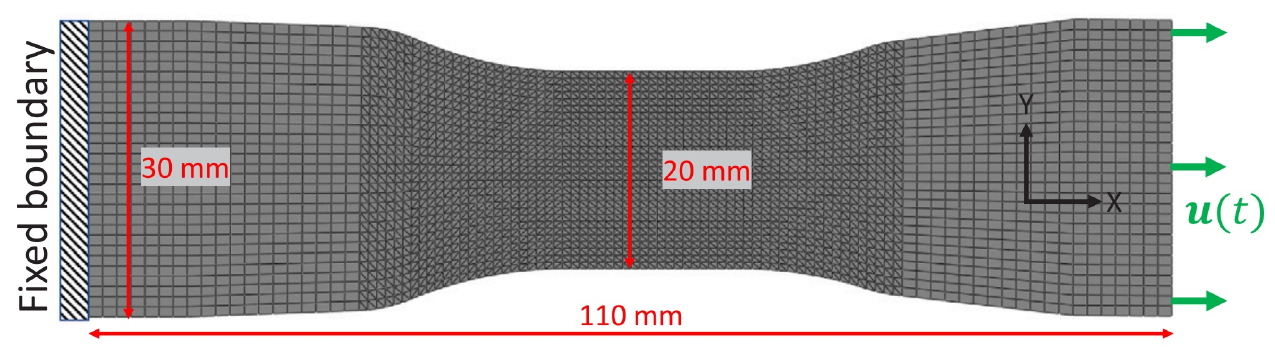}
    \caption{The dog bone specimen. }
    \label{fig:dogbone_model}
  \end{figure}

  The third example involves a dog-bone specimen subjected to sequential, time-dependent
  displacement loading, as illustrated in \cref{fig:dogbone_model}. This dataset,
  originally introduced by \citet{he2024sequential}, features the specimen fixed
  on the left boundary, while the right boundary experiences a randomly
  generated displacement sequence over time, following the same protocol as the previous
  cases. The mechanical behavior is described by an elastic-plastic constitutive
  model with linear isotropic hardening. Comprehensive details regarding the material
  parameters, finite element modeling, and data generation can be found in \citet{he2024sequential}.
  The dataset comprises 15,000 samples, each containing a loading profile, the finnal
  state of von Mises stress, and plastic equivalent strain (PEEQ). For model evaluation,
  3,200 samples are used for training and 800 for testing to assess the
  generalization performance of S-NOT.

  \section{Results}
  \label{sec:results}

  In this section, we present the prediction results for the three benchmark datasets
  introduced in \cref{sec:examples_datasets}. To evaluate model performance, we use
  the relative $L_{2}$ error for stress predictions, defined as
  \begin{equation}
    \label{eq:mean_l2}L^{2}= \frac{1}{N_{P}}\sum_{i=1}^{N_P}\frac{\left\| \boldsymbol{S}_{i}^{\text{true}}-
    \boldsymbol{S}_{i}^{\text{predict}}\right\|_{2}}{\left\| \boldsymbol{S}_{i}^{\text{true}}\right\|_{2}}
  \end{equation}
  where $\boldsymbol{S}_{i}^{\text{true}}$ and
  $\boldsymbol{S}_{i}^{\text{predict}}$ denote the true and predicted full-field
  solutions for the $i$-th test sample, and $\|\cdot\|_{2}$ is the Euclidean norm.

  For the PEEQ predictions, we use the mean absolute error (MAE) as the
  evaluation metric, defined as
  \begin{equation}
    \label{eq:mean_mae}\mathrm{MAE}= \frac{1}{N_{P}}\sum_{i=1}^{N_P}\left| \boldsymbol
    {S}_{i}^{\text{true}}- \boldsymbol{S}_{i}^{\text{predict}}\right|
  \end{equation}

  The performance of S-NOT is compared against S-DON, with a summary provided in
  \cref{tab:results} and detailed results discussed in the following subsections.

  \cref{tab:comp_efficiency} compares the computational efficiency and model size
  of S-NOT and S-DON. S-NOT offers similar inference speed and parameter count
  to S-DON, while delivering improved prediction accuracy.

  \begin{table}[hbt]
    \centering
    \caption{Comparison of S-NOT and S-DON performance on the three benchmark
    datasets. Reported values are mean errors over all test samples, with
    standard deviation in parentheses. '--' indicates unavailable data.}
    \begin{tabular}{c|cc|cc|cc}
      \hline
                           & \multicolumn{2}{c|}{Stress $L_{2}$ error [\%]} & \multicolumn{2}{c|}{PEEQ MAE [$\times10^{-4}$]} & \multicolumn{2}{c}{Temperature $L_{2}$ error [\%]} \\
      \hline
      Method               & S-DON                                          & S-NOT                                           & S-DON                                             & S-NOT       & S-DON        & S-NOT       \\
      \hline
      Steel Solidification & 18.1 (23.2)                                    & 4.3 (15.3)                                      & --                                                & --          & 0.091 (0.49) & 0.04 (0.52) \\
      3D Lug               & 11.6 (36.5)                                    & 5.31 (15.1)                                     & 1.4 (4.74)                                        & 0.76 (3.67) & --           & --          \\
      Dog-bone             & 2.01 (6.19)                                    & 1.13 (3.59)                                     & 0.79 (2.37)                                       & 0.27 (1.5)  & --           & --          \\
      \hline
    \end{tabular}
    \label{tab:results}
  \end{table}

  \begin{table}[hbt]
    \centering
    \caption{Comparison of computational efficiency and model size for S-NOT, S-DON,
    and FEM across the three benchmark problems. Reported are the number of trainable
    parameters and average inference time per sample (GPU), alongside reference
    FEM simulation time (CPU).}
    \begin{tabular}{c|cc|cc|c}
      \hline
                           & \multicolumn{2}{c|}{Number of param} & \multicolumn{2}{c|}{Inference time (GPU) [s/sample]} & \multirow{2}{8em}{FEM simulation (CPU) [s/sample]} \\
      \cline{1-5} Method   & S-DON                                & S-NOT                                                & S-DON                                             & S-NOT   &     \\
      \hline
      Steel Solidification & 772,674                              & 805,592                                              & 7.33e-4                                           & 1.1e-3  & 333 \\
      3D Lug               & 2,007,426                            & 1,138,689                                            & 4.7e-3                                            & 2.87e-3 & 80  \\
      Dog-bone             & 771,906                              & 7,90952                                              & 9e-4                                              & 1.2e-3  & 20  \\
      \hline
    \end{tabular}
    \label{tab:comp_efficiency}
  \end{table}

  \subsection{Multiphysics Steel Solidification}

  \begin{figure}[h]
    \centering
    \includegraphics[width=\textwidth]{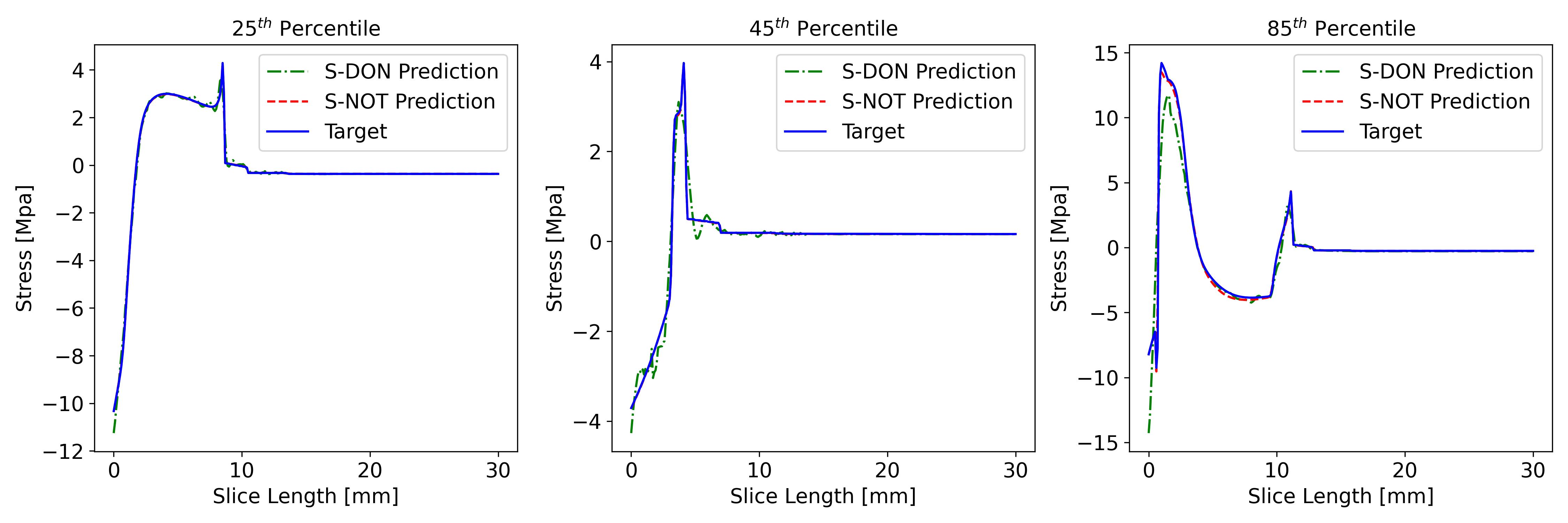}
    \caption{Stress Comparison, Multiphysics Solidification.}
    \label{fig:slice_stress_comparison}
  \end{figure}

  \begin{figure}[h]
    \centering
    \includegraphics[width=\textwidth]{
      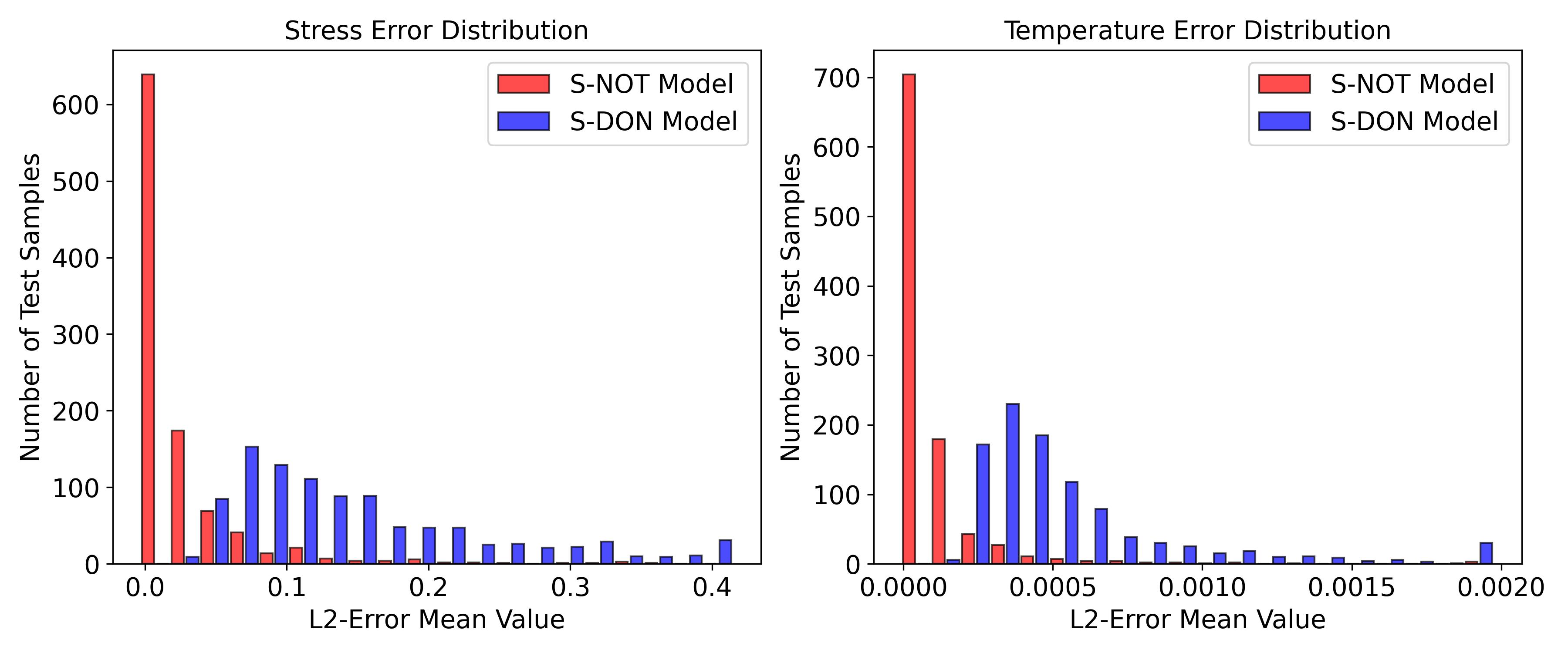
    }
    \caption{Test L2 errors Distribution (histograms) of the Multiphysics Slice.}
    \label{fig:slice_histogram}
  \end{figure}

  Both S-DON and S-NOT accurately predict the temperature distributions, which are
  relatively smooth, as indicated by the low mean relative $L_{2}$ norm error
  across all test samples (\cref{eq:mean_l2}). In contrast, stress distributions
  are more challenging due to the influence of three constitutive models (austenite,
  delta-ferrite, and mushy/liquid), temperature-dependent thermal stresses, and
  the complex, time-varying displacement history applied to the cooled surface.
  These factors lead to highly irregular lateral stress profiles throughout the
  slice domain, making precise prediction difficult. S-NOT achieves
  substantially higher accuracy, with a mean stress $L_{2}$ error of \textbf{4.3}\%,
  compared to \textbf{18.1}\% for S-DON, and a mean temperature $L_{2}$ error of
  \textbf{0.04}\% versus \textbf{0.091}\% for S-DON, as shown in
  \cref{tab:results}.

  \cref{fig:slice_stress_comparison} shows predicted and reference stress values
  at the mold exit along the slice domain for test samples at the 25th, 45th,
  and 85th percentiles (ranked by $L_{2}$ error). The difference between models is
  especially pronounced in the 85th percentile sample, where S-DON deviates
  significantly from both the target and S-NOT predictions, particularly at the chilled
  surface (0 mm) and within the solid phase. Accurate prediction of lateral
  stress at the chilled surfaces is critical for forecasting failure mechanisms
  in steel casting, such as hot tearing, highlighting the importance of S-NOT's
  improved performance.

  The test error histograms for both temperature and stress in
  \cref{fig:slice_histogram} further illustrate the superior accuracy of S-NOT
  over S-DON, where S-DON exhibits a longer tail of higher error values.

  \subsection{3D-LUG}
  \label{sec:3Dlug}

  For the 3D-LUG example, model accuracy is evaluated using the relative $L_{2}$
  error for stress and the mean absolute error (MAE) for PEEQ. S-NOT achieves a
  relative $L_{2}$ error of 5.31\% for stress, outperforming S-DON's 11.6\%. For
  PEEQ, S-NOT attains an MAE of $7.62\times 10^{-5}$, which is notably lower
  than S-DON's $1. 4\times 10^{-4}$.

  \begin{figure}[!h]
    \centering
    \includegraphics[width=\textwidth]{
      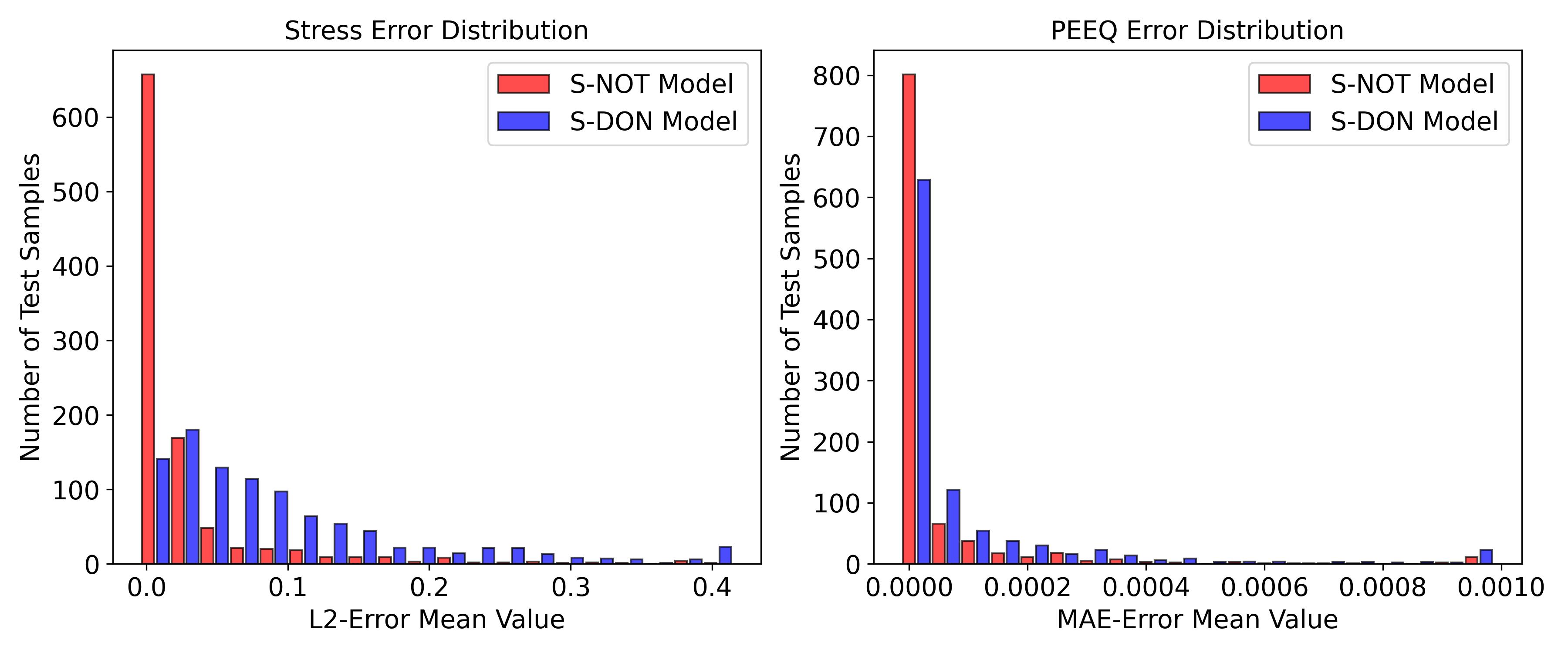
    }
    \caption{ test L2 errors Distribution (histograms) of the 3D LUG.}
    \label{fig:3Dlug_histogram}
  \end{figure}

  \begin{figure}[!h]
    \centering
    \includegraphics[width=\textwidth]{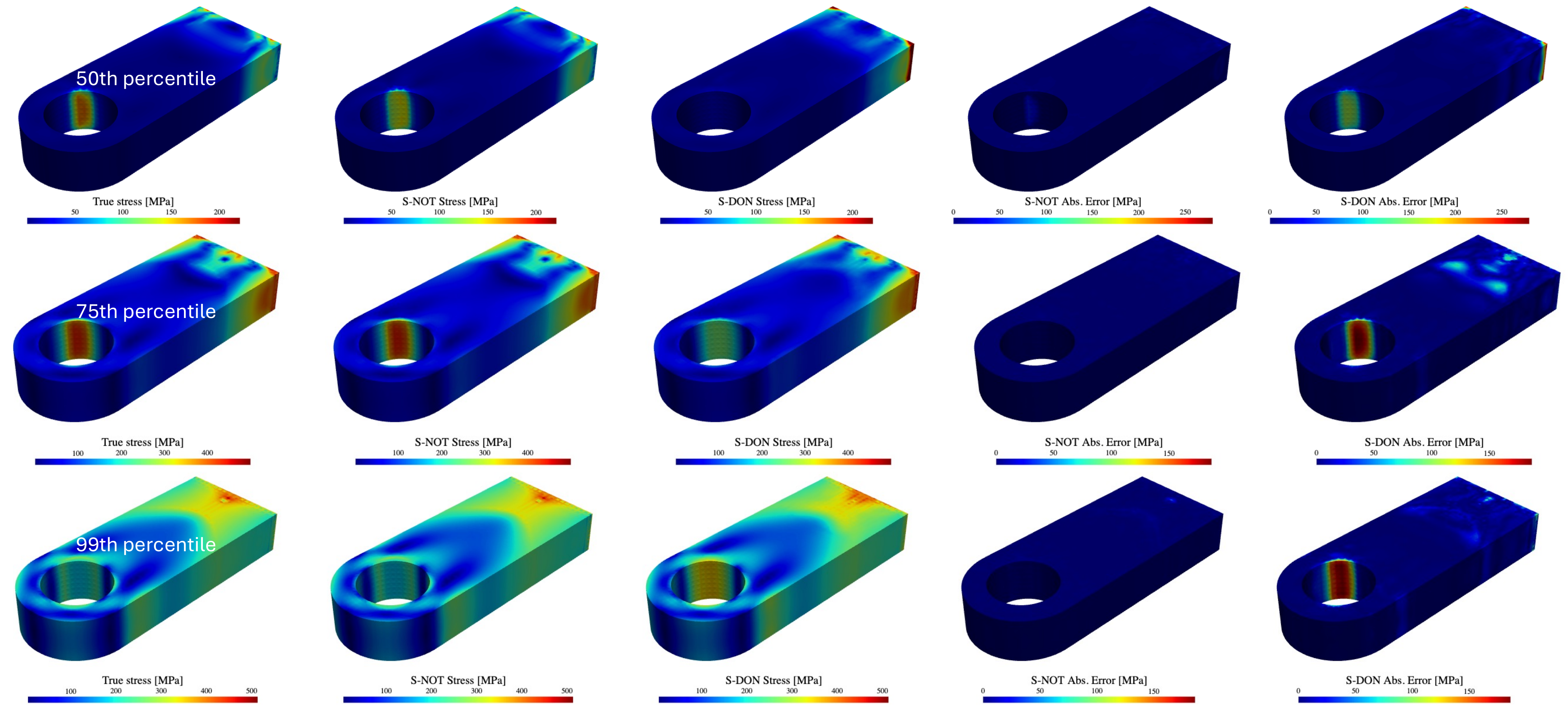}
    \caption{Stress comparison for the 3D LUG example: True vs S-NOT vs S-DON.
    Each row corresponds to a test sample at the 50th, 75th, and 99th
    percentiles (by relative $L_{2}$ error). Columns show: (1) true stress, (2) S-NOT
    prediction, (3) S-DON prediction, (4) error between S-NOT and true, and (5)
    error between S-DON and true.}
    \label{fig:3DlugStress}
  \end{figure}

  \begin{figure}[!h]
    \centering
    \includegraphics[width=\textwidth]{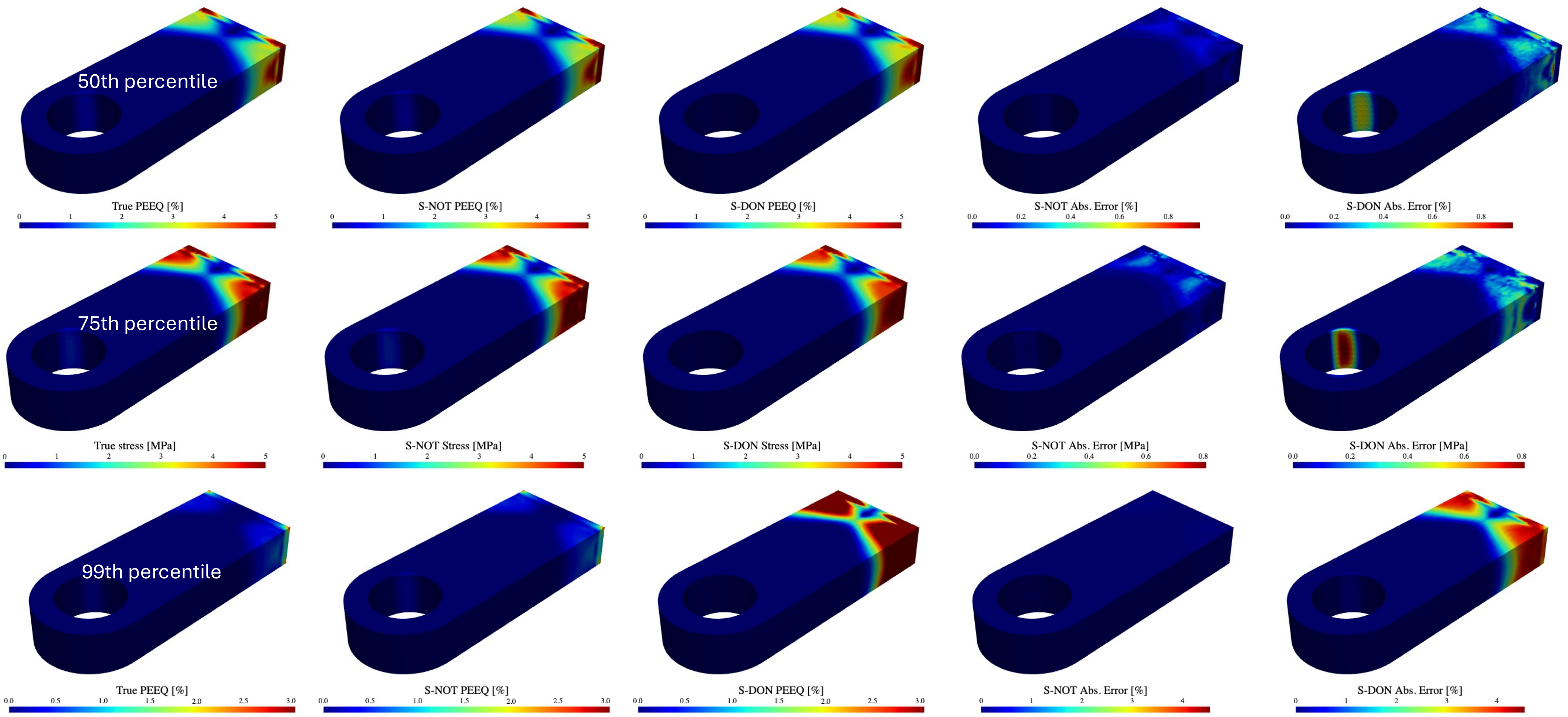}
    \caption{PEEQ comparison for the 3D LUG example: True vs S-NOT vs S-DON.
    Rows correspond to the 50th, 75th, and 99th percentile test samples (by relative
    $L_{2}$ error). Columns show: (1) true PEEQ, (2) S-NOT prediction, (3) S-DON
    prediction, (4) error between S-NOT and true, and (5) error between S-DON
    and true.}
    \label{fig:3DlugStrain}
  \end{figure}

  The histograms of test error distributions for both PEEQ and stress in \cref{fig:3Dlug_histogram}
  highlight the superior accuracy of S-NOT compared to S-DON, particularly for
  stress predictions, where S-DON exhibits a longer tail of higher error values (blue
  columns), and less samples with high accuracy compared to S-NOT (red columns).

  \cref{fig:3DlugStress} presents a comparison of predicted and reference stress
  values for test samples at the 50th, 75th, and 99th percentiles (by relative $L
  _{2}$ error). The difference between the models is especially pronounced in these
  cases, with S-DON deviating significantly from the target in the hole region, while
  S-NOT predictions remain much closer to the reference across the entire domain.
  A similar trend is observed for PEEQ in \cref{fig:3DlugStrain}, where S-NOT consistently
  achieves lower absolute errors than S-DON.

  \subsection{Dog-bone}

  For the dog-bone example, model performance is evaluated using the relative $L_{2}$
  error for stress and the mean absolute error (MAE) for PEEQ. S-NOT achieves a
  relative $L_{2}$ error of 1.13\% for stress, outperforming S-DON's 2.01\%. For
  PEEQ, S-NOT attains an MAE of $2.67\times 10^{-5}$, which is notably lower
  than S-DON's $7.85\times 10^{-5}$.

  \begin{figure}[!h]
    \centering
    \includegraphics[width=\textwidth]{
      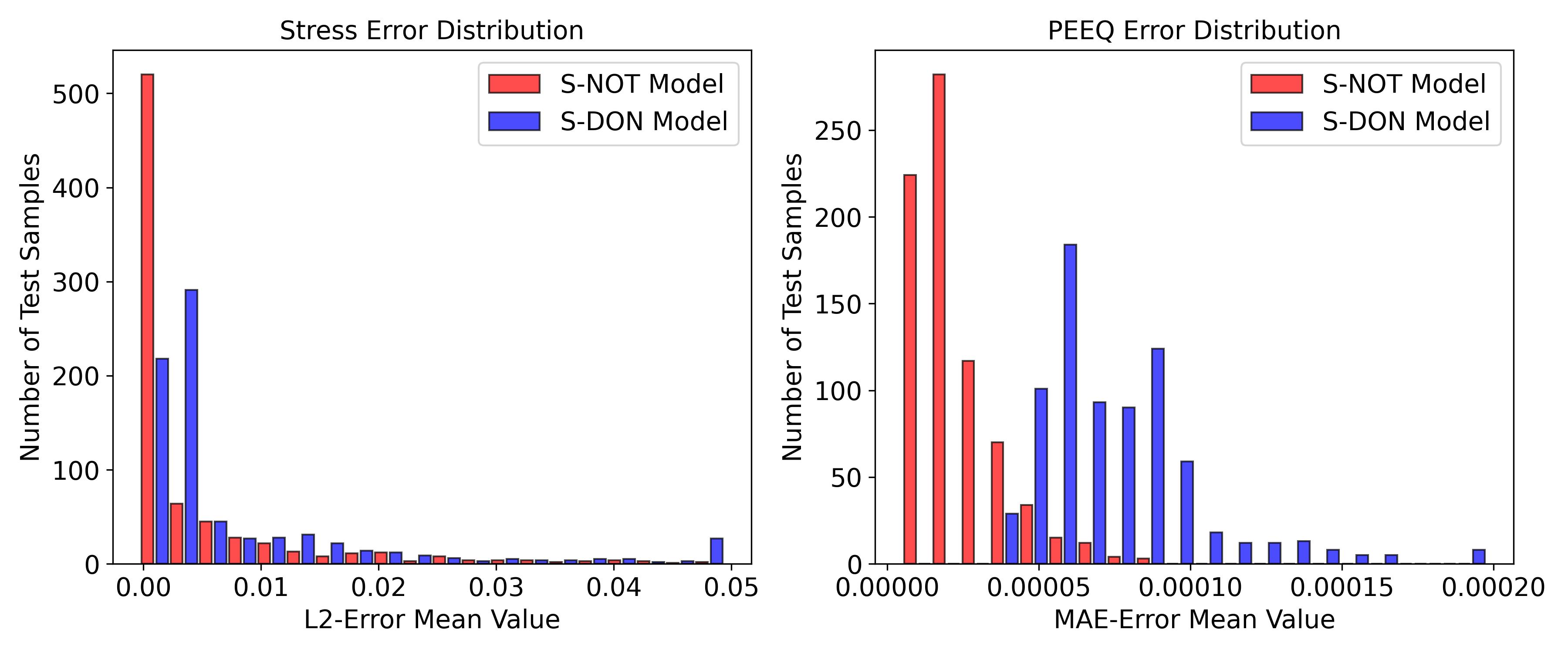
    }
    \caption{ test L2 errors Distribution (histograms) of the Dog-bone.}
    \label{fig:dogbone_histogram}
  \end{figure}

  \begin{figure}[!h]
    \centering
    \includegraphics[width=\textwidth]{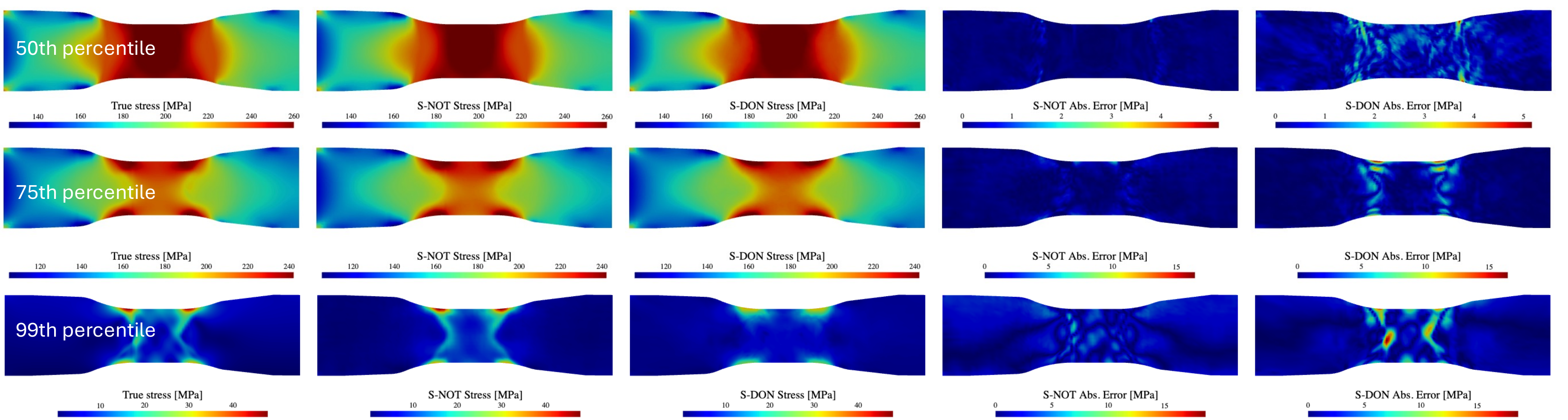}
    \caption{Stress comparison for the dog-bone example: True vs S-NOT vs S-DON.
    Rows correspond to the 50th, 75th, and 99th percentile test samples (by relative
    $L_{2}$ error). Columns show: (1) true stress, (2) S-NOT prediction, (3) S-DON
    prediction, (4) error between S-NOT and true, and (5) error between S-DON
    and true.}
    \label{fig:dogbone_stress}
  \end{figure}

  \begin{figure}[!h]
    \centering
    \includegraphics[width=\textwidth]{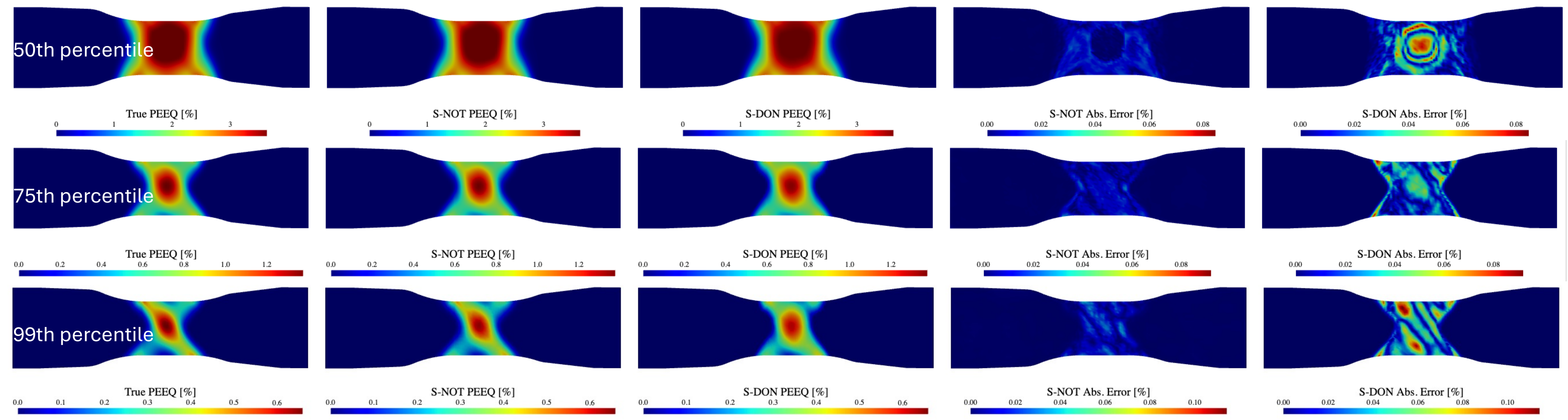}
    \caption{PEEQ comparison for the dog-bone example: True vs S-NOT vs S-DON.
    Rows correspond to the 50th, 75th, and 99th percentile test samples (by relative
    $L_{2}$ error). Columns show: (1) true PEEQ, (2) S-NOT prediction, (3) S-DON
    prediction, (4) error between S-NOT and true, and (5) error between S-DON
    and true.}
    \label{fig:dogbone_strain}
  \end{figure}

  The histogram of test error distributions for both PEEQ and stress in \cref{fig:dogbone_histogram}
  highlights the improved accuracy of S-NOT over S-DON, particularly for PEEQ
  predictions, where the majority of S-NOT results (red columns) outperform even
  the best S-DON predictions (blue columns).

  \cref{fig:dogbone_stress} presents a comparison of predicted and reference stress
  values for test samples at the 50th, 75th, and 99th percentiles (by relative
  $L_{2}$ error). The difference between the models is especially pronounced in
  these cases, with S-DON deviating significantly from the target in the critical
  regions, while S-NOT predictions remain much closer to the reference across
  the entire domain. A similar trend is observed for PEEQ in \cref{fig:dogbone_strain},
  where S-NOT consistently achieves lower absolute errors than S-DON.

  \section{Conclusions}
  \label{sec:conclusion}

  In this work, we presented the Sequential Neural Operator Transformer (S-NOT),
  a novel deep learning framework for efficiently solving time and path-dependent
  nonlinear partial differential equations (PDEs). By combining gated recurrent
  units (GRUs) for sequential encoding with transformer-based attention
  mechanisms, S-NOT captures complex temporal and spatial dependencies in physical
  systems. The use of cross-attention enables the model to selectively aggregate
  relevant information for each query point, addressing the limitations of dot product-based
  approaches such as Sequential DeepONet (S-DON).

  Benchmarking S-NOT on three highly nonlinear datasets originating from real-world
  simulations of multiphysics steel solidification, a 3D lug specimen, and a dog-bone
  specimen under sequential loadings, consistently demonstrated improved
  prediction accuracy and generalization compared to S-DON, particularly among outlier
  test samples with a higher than 90th percentile error, where S-DeepONet’s
  accuracy regularly dropped. These results indicate that S-NOT is a robust and an
  effective surrogate model for enabling scientific and engineering workflows that
  require frequent, rapid, and accurate full-field solutions of time and path-dependent
  PDEs, including uncertainty quantification and sensitivity analysis, optimizations,
  inverse designs, online controls, and digital twins.

  \section{Data availability}
  The dataset is available on Zenodo and the trained models are available at the
  GitHub repository \url{https://github.com/QibangLiu/SNOT}

  \section{Code availability}
  The codes for training and inference are available at the GitHub repository at
  \url{https://github.com/QibangLiu/SNOT}

  \section{Acknowledgements}
  The authors would like to thank the National Center for Supercomputing
  Applications (NCSA) at the University of Illinois, and particularly its Research
  Computing Directorate, Industry Program, and Center for Artificial Intelligence
  Innovation (CAII) for support and hardware resources. This research used both
  the DeltaAI advanced computing and data resource, which is supported by the National
  Science Foundation (award OAC 2320345) and the State of Illinois, and the
  Delta advanced computing and data resource which is supported by the National
  Science Foundation (award OAC 2005572) and the State of Illinois. Delta and DeltaAI
  are joint efforts of the University of Illinois Urbana-Champaign and its National
  Center for Supercomputing Applications.

  \section{Author contributions statement}

  \textbf{Q. Liu} and \textbf{S. Koric} jointly conceptualized the study, designed
  the methodology, implemented the models, performed the analyses, and
  contributed to writing, reviewing, and editing the manuscript. \textbf{S. Koric}
  was responsible for dataset preparation.

  \section{Competing interests statement}
  The authors declare no competing financial or non-financial interests.

  %% Loading bibliography style file
  \bibliographystyle{elsarticle-num-names}

  % \bibliographystyle{cas-model2-names}

  %\bibliographystyle{unsrt}
  % Loading bibliography database
  \bibliography{references}

  %\vskip3pt
\end{document}